\documentstyle[graphicx]{mn}

\newif\ifAMStwofonts
\AMStwofontstrue

\def\pg{{PG1211+143}}

\def\xmm{{\it XMM-Newton}}

\def\et{{et al.\ }}

\def\asca{{\it ASCA}}

\newcommand{\ls}{\mathrel{\hbox{\rlap{\hbox{\lower4pt\hbox{$\sim$}}}\hbox{$<$}}}}
\newcommand{\gs}{\mathrel{\hbox{\rlap{\hbox{\lower4pt\hbox{$\sim$}}}\hbox{$>$}}}}

\def\arcs{{\hbox{$^{\prime\prime}$}}}

\def\Msun{\hbox{$\rm ~M_{\odot}$}}

\def\H0{{\rm ~km~s^{-1}~Mpc^{-1}}}

\def\msun{M_{\rm \odot}}

\def\msun{{\rm M_{\odot}}}

\def\et{{et al.}}

\title[A high velocity outflow from \pg]
        {Is the X-ray spectrum of the narrow emission line QSO \pg\ defined by its energetic outflow?}
\author[K.A.Pounds \et]
        {K.A.Pounds,$^{1}$
	J.N.Reeves,$^{2}$\\
$^1$ Department of Physics and Astronomy, University of Leicester,
Leicester, LE1 7RH, UK\\
$^2$ Laboratory for High Energy Astrophysics, NASA Goddard Space Flight Center, Greenbelt, MD 20771, USA\\}
\date{Accepted ; Submitted }
\pagerange{\pageref{firstpage}--\pageref{lastpage}}
\pubyear{2006}
\begin{document}
\maketitle
\label{firstpage}

\begin{abstract} 
An \xmm\ observation of the bright QSO \pg\ in 2001 revealed a blue-shifted absorption line spectrum indicative of a high velocity radial outflow of highly
ionised gas.  Unless highly collimated, the outflow mass rate was shown to be comparable to the accretion rate, with mechanical energy a significant
fraction of the bolometric luminosity. Analysis of the full \xmm\ data set now allows the wider effects of that energetic outflow  to be explored. We
find that absorption and re-emission of the primary continuum flux in the ionised outflow, together with a second, less strongly absorbed, continuum
component can explain the strong `soft excess' in \pg\ without the extreme velocity `smearing' in conflict with observed absorption line widths.
Previously  unpublished data from a second \xmm\ observation of \pg\ is shown to be consistent with the new spectral model, 
finding that the additional continuum component dominates the spectral variability. We speculate that this
variable continuum component is powered by the high velocity outflow. 
\end{abstract}


\section{Introduction}
 
The luminous QSO \pg\ has been observed twice by \xmm, in 2001 and 2004. Analysis of the  EPIC and RGS spectra from the first observation
provided evidence for a highly ionised outflow with a velocity of $\sim$0.09$\pm$0.01c (Pounds \et\ 2003; hereafter P03), though this high velocity was
contested following a careful ion-by-ion modelling of the RGS data (Kaspi and Behar 2006). However, a re-examination of the 2001 EPIC
data (Pounds and Page 2006) resolved additional absorption lines in the EPIC MOS spectrum, strengthening the initial claim, and yielding a preferred velocity of
v$\sim$0.14c. Confirmation of the high velocity outflow is important since, as noted in P03, the mechanical energy in the flow, if not highly collimated, is a
significant fraction of the bolometric luminosity of \pg\ and could be typical of AGN accreting near the Eddington rate (King and Pounds 2003), while also
providing an example of the feedback required by the linked growth of SMBH in AGN with their host galaxy (King 2004).

P03 also found a strong `soft excess'
over a 2--10 keV power law fit of photon index $\Gamma$$\sim$1.8, a typical value for type 1 AGN, and a broad Fe K emission line. Subsequently, the 2001 EPIC
spectrum of \pg\ has been used by Gierlinski and Done (2004), and more recently by Schurch and Done (2006), to demonstrate how strong absorption of the intrinsic
X-ray continuum in a `velocity-smeared', high  column, ionised gas could provide an alternative explanation (to Comptonisation) for the strong soft
excess generally seen in type 1 AGN (Wilkes and Elvis 1987, Turner and Pounds 1989). A more general study by Chevallier \et\ (2006) also considered an ionised
reflection origin of the soft excess in AGN (e.g.Crummy \et\ 2006) and concluded that absorption was the more likely  cause of a strong soft excess (as in \pg). 

In the present paper we use the full \xmm\ data set from both observations of \pg\ to re-examine the broad-band X-ray spectrum of \pg, to better understand the
structure and dynamics of the outflow, and and thereby to assess the 
conjecture that the massive and energetic outflow substantially defines the emerging spectrum.

We assume a redshift for \pg\ of $z=0.0809$ (Marziani \et\ 1996).

\section{Observation and data reduction}

\pg\ was observed by \xmm\ on 2001 June 15 for $\sim$53 ks, and again on 2004 June 21 for $\sim$57 ks. In this paper we mainly use the high signal-to-noise 
data from the EPIC
cameras (Str\"{u}der \et\ 2001; Turner \et\ 2001), but with a crucial constraint on high resolution spectral features provided by the simultaneous RGS
data (den Herder \et\ 2001). All X-ray data were first screened with the XMM SAS v6.5 software and events selected corresponding to patterns 0-4 (single and
double pixel events) for the pn camera and patterns 0-12 for the MOS cameras. We extracted EPIC source counts within a  circular region of
45\arcs\ radius defined around the centroid position of \pg, with the background being taken from a similar region, offset from but close to the target source.
After removal of data during periods of high background the effective exposures for the pn camera were $\sim$49.5 ks in 2001 and $\sim$25.2 ks in 2004,
with corresponding exposures of $\sim$103 ks and $\sim$81 ks for the combined MOS cameras. Individual spectra were then binned to a minimum of 20 counts
per bin, to facilitate use of the $\chi^2$ minimalisation technique in spectral fitting. 

Spectral fitting was based on the Xspec package (Arnaud 1996), version 11.3.  All spectral fits include absorption due to the line-of-sight Galactic
column of $N_{H}=2.85\times10^{20}\rm{cm}^{-2}$  (Murphy \et\ 1996) and errors are quoted at  the 90\% confidence level ($\Delta \chi^{2}=2.7$ for one
interesting parameter).

\section{Visual examination of the EPIC data} 
  
Figure 1 displays the pn spectra of \pg\ from the 2001 and 2004 observations in a way that has become conventional, by first  fitting a power law over the
2--10 keV band and then  extrapolating this fit to lower energies, revealing the `soft excess' typical of many type 1 AGN.  This approach, followed by P03
in their initial analysis of the 2001 pn camera data of \pg, yielded a power law fit over the 2--10 keV band with a `canonical' photon index of
$\Gamma$$\sim$1.79, together with an Fe K emission line (apparently with a broad red wing), and a strong soft excess parameterised by a black body
spectrum of kT$\sim$0.1 keV. The same approach for the 2004 pn data of \pg\ yields a steeper power law ($\Gamma$$\sim$1.85) and a less strong
`soft excess' (figure 1).

\begin{figure}                                                          
\centering                                                              
\includegraphics[width=6cm, angle=270]{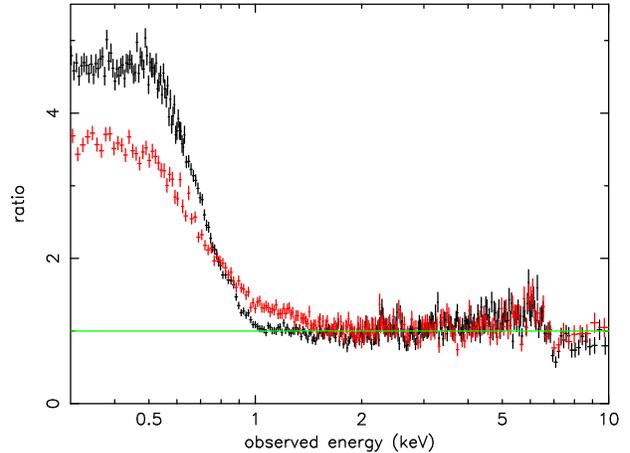}                     
\caption                                                                
{Comparison of the X-ray spectra of \pg\ displayed in the conventional way, showing the strong soft excess sitting above a `canonical' power law, 
for the pn data from the 2001 (black) and 2004 (red) observations. On this view it appears that the soft excess in 2004 is weaker but `hotter'}
\end{figure}

An alternative comparison of the 2001 and 2004 spectra is shown in figure 2, which plots the unmodelled, background-subtracted data. Shown in this way, we
see the spectral change is actually an increase in flux (from 2001 to 2004) between $\sim$0.7--2 keV, and a (less obvious) decrease at $\sim$0.5--0.7 
keV. Interestingly, the 2001 and 2004 spectra are essentially identical above $\sim$3 keV. We choose to show the MOS data in figure 2 since, while the
broad flux profile is the same, the higher energy resolution of the MOS data shows additional low energy structure in the 2001 spectrum.

\begin{figure}                                                                              
\centering                                                              
\includegraphics[width=6cm, angle=270]{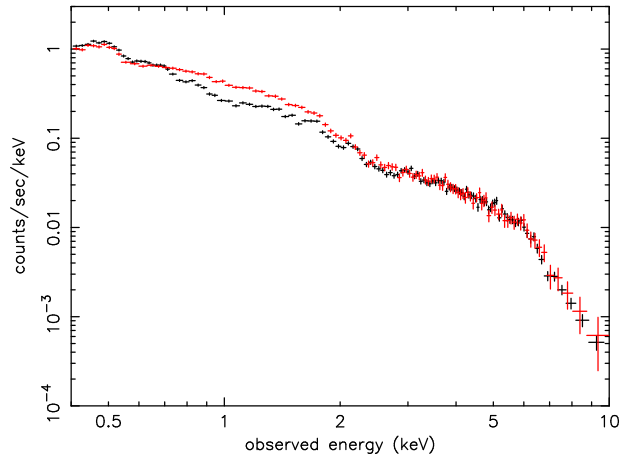} 
\caption                                                                
{Direct comparison of the background-subtracted MOS data from the 2001 (black) and 2004 (red) observations shows the spectral change to be due to an 
increase in flux between $\sim$0.7--2 keV, and a less obvious flux decrease at $\sim$0.5--0.7 keV}
\end{figure}

To highlight the broad spectral changes between the 2001 and 2004 data we then directly compared the two data sets, noting, for example, that if
absorption is having a controlling effect on the emerging spectrum, it should show up in the way in which individual spectra differ. As variable
absorption (multiplicative) spectral components are best seen in the ratio of data sets, we therefore re-binned the EPIC data from 2001 to a minimum of
200 counts per channel, in order to show any broad features more clearly, and then re-binned the 2004 data to have identical energy channels. Dividing
the first data set by the second then gave the spectral ratio plot of figure 3. The most obvious feature is a broad flux deficit over
the  $\sim$0.7-2 keV band, the shape and  positioning of which is strongly suggestive of differential absorption by ionised gas, with
enhanced absorption in the 2001 spectrum. 

Figure 4 shows the `difference spectrum', obtained by subtracting the 2004 (background-subtracted) data set from that of 2001. The data have again been
re-binned to a minimum of 200 counts per bin. While above $\sim$2 keV  the difference spectrum reflects the very similar hard X-ray fluxes, in addition to a
lower flux at $\sim$0.7-2 keV we see a significant excess peaking at $\sim$0.5 keV in the 2001 data.

\begin{figure}                                                          
\centering                                                                                                                           
\includegraphics[width=4.7cm, angle=270]{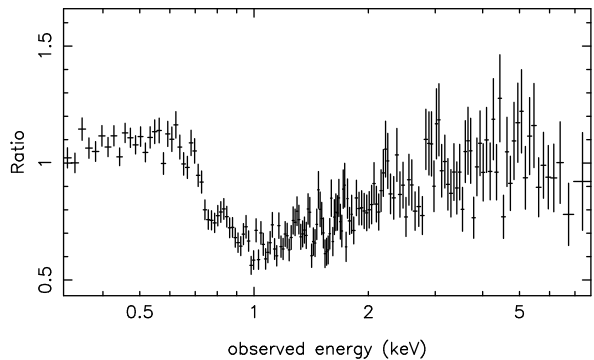}
\caption                                                                
{The MOS spectral ratio obtained by dividing the 2001 count rate spectrum by that obtained in 2004. The most obvious feature is a broad deficit at $\sim$0.7-2 keV
indicative of variable photoionised absorption}      
\end{figure}

\begin{figure}                                                          
\centering                                                              
\includegraphics[width=4.7cm, angle=270]{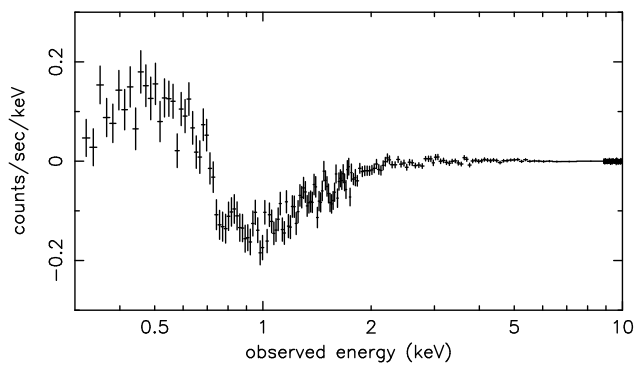}                                         
\caption                                                                
{The MOS difference spectrum, obtained by subtracting the 2004 spectrum from that of 2001, shows the near-identical hard X-ray spectra, with a lower flux at $\sim$0.7-2 keV
and comparable higher flux peaking at $\sim$0.5-0.6 keV in the 2001 observation}      
\end{figure}

In summary, a direct comparison of the 2001 and 2004 MOS data (comparison of the pn data shows the same features) suggests a decrease in continuum 
absorption is the  main cause of the spectral change in \pg\ from 2001 to 2004, together with a decrease in an emission component specific to the soft
X-ray band. Implicitly, both results support the contention that reprocessing in ionised gas is a primary factor in determining the form of the
observed broad band X-ray spectrum. An important constraint on the nature of the variable absorption is the similarity in the ionising flux during the 2001 and 2004
observations indicated by the near-identical spectra above $\sim$3 keV, suggesting a change in covering factor rather than ionisation parameter.

\section{The case for a fresh approach to modelling the X-ray spectrum of \pg} 

Accumulation of high quality, broad-band X-ray spectra of AGN, particularly from\xmm\ EPIC, has raised doubts over the the physical
reality of the  soft excess and also the ubiquity of the relativistic Fe K emission line suggested by earlier \asca\ observations.  In particular, the
conventional interpretation of the soft excess in type 1 AGN, as arising by Comptonisation of cooler disc photons, has been questioned by Gierlinski and
Done (2004).

Those authors studied 26 AGN from the PG sample observed with XMM, fitting the X-ray spectra with a double Comptonisation model. They found
that while  Comptonisation of the accretion disc emission provided a good statistical fit to the soft excess below 2 keV in every case, the temperature of 
the cool
Comptonising region was  remarkably constant over the sample, despite that sample having a large range in mass and luminosity, and hence in disc
temperature, and in the ratio of power released  in the hot plasma to that from the disc. Given these difficulties, Gierinski and Done (2004) proposed
that the soft excess was an artefact of ionised absorption in a sub-relativistic wind above the inner accretion disc. In order to explain the generally smooth form of the
soft excess the absorbing gas was required to have a complex velocity structure. Subsequently, Schurch and Done (2006) have explored this `smeared
absorber' model in more detail, showing - in particular - that by including re-emission from the ionised gas the strong soft excess in \pg\ could be
reproduced. 

Chevallier \et\ (2006) have also examined alternative origins of the soft excess in AGN, including enhanced soft X-ray emission from an ionised accretion
disk (e.g. Crummy \et\ 2006). Chevallier \et\ found that ionised absorption in a high speed outflow, or an inhomogeneous accretion flow could
best explain some AGN X-ray spectra, particularly those with very strong soft excesses (as \pg), while only weak excesses can be modelled by reflection
unless the primary continuum is strongly depressed (Fabian \et\ 2002). However, they concluded that simple `smeared' absorption
models would require `fine tuning' to constrain the depth of the trough near 1 keV and thereby be generally applicable.

A second spectral feature whose interpretation has been questioned recently is the broad wing seen to the low energy  side of the Fe K
fluorescent emission line in a number of bright AGN. While widely considered to be due to Doppler and strong gravity broadening of radiation 
reflected off the innermost accretion disc (eg Fabian \et\ 2000), an alternative explanation as (again) an artefact of absorption has been
proposed in specific cases (eg Inoue and Matsumoto 2003, Kinkhabwala 2003). 

Clarifying the nature of such significant features in the X-ray spectra of AGN is important to explore the accretion process, outflows of ejected matter, and ultimately the
properties of the central black hole. Its strong soft excess, evidence for a massive and energetic outflow, and high accretion ratio make \pg\ a particularly
interesting object by which to explore an alternative approach to deconvolving the broad band X-ray spectrum.

\section{Developing a new spectral model for the 2001 observation of \pg}

We begin the fresh approach by noting that the intrinsic continuum over the broad X-ray band now visible to \xmm\ might not be adequately described by 
a simple power law derived
by fitting to the data over the $\sim$2--10 keV band. Recent studies of spectral variability in MCG-6-30-15 (Vaughan and
Fabian 2004) and 1H0419-577 (Pounds \et\ 2004), among the most  detailed broad-band spectral analyses to date, have provided clear evidence of a 
variable power law component of slope significantly  steeper than the `canonical value' of $\Gamma$$\sim$1.9 (Nandra and Pounds 1994). That finding is important in the present
context since, if absorption is to create the impression of a strong `soft excess',  then the underlying continuum must be much steeper than it appears
above $\sim$2 keV. To allow for that possibility, while retaining a harder `primary' continuum in the range predicted by the disc/corona
model (Haardt and Maraschi 1991), we include 2 power law components in the new model. For simplicity we will call these the primary and secondary continuum components. 

Indication from the flux ratio plot that the main spectral change between the 2001 and 2004 \xmm\ observations of \pg\ is due to a change in ionised
absorption affecting the spectrum above $\sim$0.7 keV, suggests the model should include a moderately ionised (`warm') absorber, in addition to the highly ionised 
absorber responsible for the narrow, blue-shifted absorption lines (P03, Page and Pounds 2006). Finally, we include components to model the re-emission 
from both ionised
absorbers.

In Xspec terms the prospective model is then WA(PO1*ABS1*ABS2 + PO2*ABS3*ABS4 + EM1/3 + EM2/4), where WA represents the line-of-sight Galactic column,
included in all fits, and PO1,  PO2 are the separate power law components. ABS1 and 3 represent the high ionisation absorber, differing only in column
density on the 2 continua, while ABS2 and 4 similarly represent the lower ionisation absorption. Finally, EM1/3 and EM2/4 model the re-emission from the
highly ionised and warm absorbers, respectively.  All ABS and EM components are modelled by photoionised gas using the XSTAR code (Kallman et
al 1996).  We assume a relatively high turbulent velocity of 1000 km s$^{-1}$ FWHM in order to allow substantial line opacity before saturation. Free
parameters in each XSTAR grid are the column density and ionisation parameter, with outflow (or inflow) velocities included as an adjustment to the
apparent redshift of the absorbing or emitting gas. All abundant elements from C to Fe are included with the relative abundances as a further variable
input parameter, but tied for all the XSTAR components and limited to within a factor 2 of solar. Further constraints on the model are to tie the
ionisation parameters of ABS1 and 3, and of ABS2 and 4, and also with their respective re-emission spectra.

We first evaluated the model with the pn and MOS data sets, separately, given their differing low energy responses and the higher MOS resolution, which
potentially  is important in resolving features in the soft X-ray band (Pounds and Page 2006).

\begin{figure}                                                          
\centering                                                              
\includegraphics[width=6cm, angle=270]{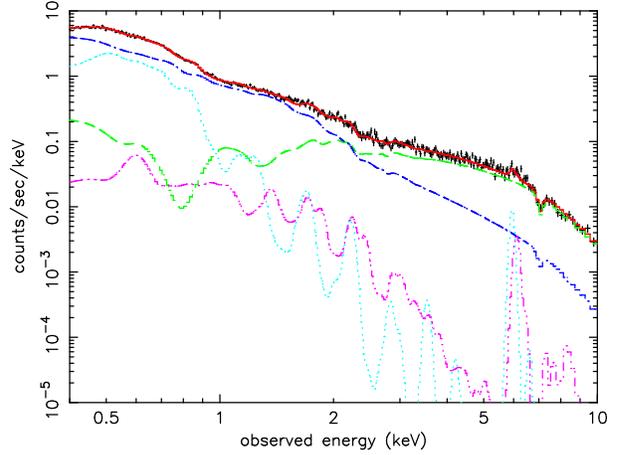}                                          
\caption                                                                
{The full spectral model (red) fitted to the 2001 pn camera data of \pg. The model components are the primary power law (green), secondary power law (dark blue), 
re-emission from the moderately ionised gas (light blue) and re-emission from the highly ionised gas (magenta)}      
\end{figure}

An excellent statistical fit was obtained for the pn data with the component parameters as listed in Table 1. The spectral components are shown together
with the total model and pn data in figure 5. Fitting the model to the 2001 MOS data yielded similar component parameters, although the statistical quality of the fit was less good. The 
fit parameters are listed in line 2 of Table 1 and the data, model and separate model components are shown in figure 6.
 
Visual examination of figures 5 and 6 clarifies the main features of the new model.

Above $\sim$3 keV the spectrum is dominated by the `primary' power law, with a
photon index $\Gamma$$\sim$2.2, at the upper end of - but consistent with - the `accepted' range for type 1 AGN spectra. At lower energies the 
secondary power law
component ($\Gamma$$\sim$3.1) is more important, while below $\sim$1 keV re-emission, particularly from the `warm' absorber, becomes significant. We note
that variability in this component between 2001 and 2004 could explain the low energy peak in the difference spectrum of figure 4.

\begin{figure}                                                          
\centering                                                              
\includegraphics[width=6cm, angle=270]{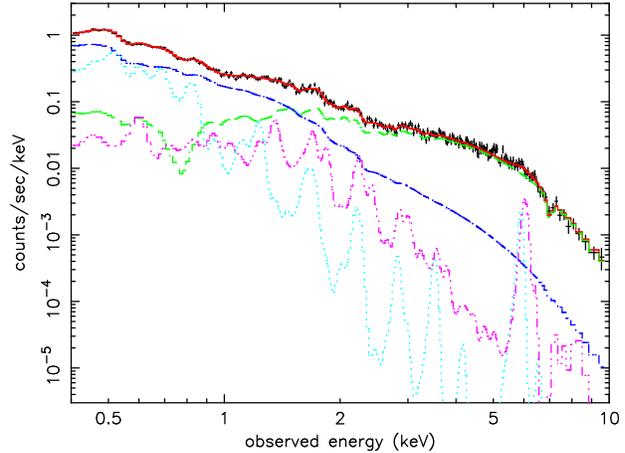}                     
\caption                                                                
{The spectral model (red) fitted to the 2001 MOS camera data of \pg. The model components are primary power law (green), additional power law (dark blue), 
re-emission from the moderately ionised gas (light blue) and re-emission from the highly ionised gas (magenta)} 
\end{figure}

Strong absorption of the primary power law is seen to be responsible for the mid-band ($\sim$2-6 keV) spectral curvature, with ionised gas
column densities of N$_{H}$$\sim$$3-5\times 10^{22}$ cm$^{-2}$. In both pn and MOS spectral fits the model shows a broad trough near $\sim$0.8 keV, due
to Fe-L absorption. Figure 3 shows a similar spectral feature in the 2001/2004 ratio plot, again indicating lower absorption in the second observation. 

The secondary power law continuum appears to suffer much less absorption, with a negligible warm absorber column and a factor $\sim$2-5 lower column density of
highly ionised absorber.

The high velocity, high ionisation absorber produces the narrow absorption lines reported in PO3 (with revised identifications in Pounds and Page
2006), while dilution by the steep power law component could explain the weakness of narrow absorption lines in the RGS data (P03, Kaspi and Behar 2006).

\begin{table*}
\centering
\caption{Parameters of the model fits to the pn, MOS and RGS data from the 2001 \xmm\ observation of \pg. Power law indices $\Gamma$$_{1}$ and 
$\Gamma$$_{2}$ refer to the primary and secondary continuum components, respectively. Both high and moderate ionisation absorbers affect each 
continuum component, with equivalent hydrogen column densities in units of $10^{22}$ cm$^{-2}$ and ionisation parameters in erg cm s$^{-1}$. Effective redshifts and 
implied velocities from the ionised gas absorption and re-emission spectra are given in the text}
\begin{tabular}{@{}lccccccccccc@{}}
\hline
Inst. & $\Gamma$$_{1}$ & $N_{H}$ & log$\xi$ & $N_{H}$ & log$\xi$  & $\Gamma$$_{1}$ & $N_{H}$ & log$\xi$ & $N_{H}$ & log$\xi$  &  $\chi^{2}$/dof \\
\hline
pn & $2.2\pm$0.1 &  5.5 & 1.43$\pm$0.03 & 3.8 & 2.9$\pm$0.1 & $3.1\pm$0.1 & 0.16 & 1.43$\pm$0.03 & 1.6 & 2.9$\pm$0.1 & 835/824\\
MOS & $2.2\pm$0.1 & 2.5 & 1.53$\pm$0.05 & 6.5 & 2.8$\pm$0.1 & $3.3\pm$0.1 &  0.1  & 1.53$\pm$0.05 & 1.2 & 2.8$\pm$0.1 &  438/367 \\
RGS & 2.2 & 3.7 & 1.4$\pm$0.1 & - & - & $3.3\pm$0.2 &  0.1 & 1.4$\pm$0.1 & 0.02 & 2.5$\pm$0.1  & 588/575 \\
\hline
\end{tabular}
\end{table*}

\subsection{Velocity structure}

The velocity structure of the photoionised gas components is of particularly interest and relies on identifying significant spectral structure in the
data. Our approach in P03 and in Pounds and Page (2006) was based on the visual identification of narrow absorption lines, backed up by modelling with
XSTAR. While that works well where the aim, as there, was to determine the velocity of a strong outflow, here we have to rely principally on modelling to
interpret the multi-component broad-band spectrum of \pg. For the highly ionised absorber, the high velocity is again constrained by the presence of
narrow absorption lines in the EPIC data, with the highly ionised XSTAR  component having an apparent blue-shift (in the observer frame) of 
$\sim$0.06$\pm$0.01, indicating an outflow velocity (in the AGN rest frame) of v$\sim$0.14$\pm$0.01c. For the warm absorber the XSTAR fits 
yield apparent
redshifts of $\sim$0$\pm$0.02 (pn) and $\sim$0.02$\pm$0.02 (MOS). Constraints on the velocity of the moderately ionised absorber are  less secure, being
derived from fitting the mid-band spectral curvature and the rather complex absorption structure below $\sim$2 keV where the ionised gas recovers transparency.
The column density and ionisation parameter also affect the spectral curvature while, at lower energies, the re-emission spectrum has
overlapping spectral structure. It does appear, however, that the moderately ionised gas is outflowing at a somewhat lower, but still high velocity.

In contrast, the emission line spectra in both pn and MOS models prefer a low mean velocity in the rest frame of the galaxy, with an apparent redshift
of the XSTAR emission spectra of $\sim$0.07$\pm$0.01, corresponding to a mean outflow velocity in the rest frame of \pg\ of 
$\sim$3000$\pm$3000 km s$^{-1}$. Visual
examination of the spectral components in figures 5 and 6 suggests the velocity sensitivity of the emission line gas derives from the strong OVII and
OVIII lines at $\sim$0.5 and $\sim$0.6 keV.

\subsection{Component luminosities, absorption and re-emission}  

Overall and component luminosities in the pn and MOS model fits are sufficiently similar to take their average values. For the 2001 observation we find an overall
luminosity L$_{0.4-10}$$\sim$$1.2\times 10^{44}$~erg s$^{-1}$, with a 2--10 keV  luminosity L$_{2-10}$$\sim$$5.0\times 10^{43}$~erg s$^{-1}$. The primary 
power law is the dominant component with L$_{0.4-10}$$\sim$$5\times 10^{43}$~erg s$^{-1}$; this increases to
$\sim$$1.3\times 10^{44}$~ergs$^{-1}$ when corrected for absorption (90 \% being in the moderately ionised absorber). The secondary power law has an observed
luminosity L$_{0.4-10}$$\sim$$6\times 10^{43}$~erg s$^{-1}$, increasing to $\sim$$7\times 10^{43}$~erg s$^{-1}$ when corrected for absorption.

Re-emission from the ionised outflow provides a strong contribution to the  observed soft X-ray flux below $\sim$1 keV, with an integrated luminosity in
the emission line spectrum L$_{em}$$\sim$$1.8\times 10^{43}$~erg s$^{-1}$ in the moderately ionised component and L$_{em}$$\sim$$1.5\times 10^{42}$~erg
s$^{-1}$ in the high ionisation component. Comparing the absorbed and re-emitted luminosities we estimate the covering factors, for an optically thin
radial flow, of
CF$\sim$0.2 and CF$\sim$0.1 for the warm and highly ionised gas components, respectively.

\section{Checking for consistency with the RGS spectrum of 2001}

A key constraint of the new spectral model is to produce a strong soft excess with no strong narrow spectral lines in absorption or
emission, although there should be weak absorption lines imposed in the soft band by the highly ionised, high velocity flow. Several such lines, at low
signal-to-noise in the RGS data, were claimed in our initial analysis reported in P03. 

Examination of figures 5 and 6 suggest the narrow absorption line spectrum would be significantly `diluted' by the dominance, at low energies, of the weakly absorbed
secondary power law continuum. However, the absence of strong, narrow emission lines in the RGS spectrum requires substantial line broadening of the
re-emission component in the EPIC model spectrum. 
Conceivably this could be a result of a high velocity flow integrated over a wide angle, which would be consistent with the estimated covering factor 
of $\sim$0.2, for a partially filled outflow. 

The high intrinsic resolution of the RGS therefore offers a valuable check on the model spectrum of \pg\ over the soft X-ray band, even if the emission lines
are strongly broadened. Testing the model against the RGS  data from the 2001 observation initially gave an unacceptable ($\chi^{2}$  per dof = 845/636), 
with the steep power law ($\Gamma$$\sim$3) dominant. Examination of the data:model residuals showed this to be due to substantial low energy structure, 
too broad to be fitted by the narrow line input spectrum of the XSTAR model. We therefore added a gaussian smoothing factor to the emission line component in the 
model (gsmooth in Xspec). The broadened emission lines now increased in strength to match the observed spectral structure, with the power law component
becoming correspondingly weaker, yielding an excellent fit ($\chi^{2}$  per dof = 649/635 over the 0.35--1.5 keV band). Both unsmoothed and smoothed
spectral fits are reproduced in figure 7, where the soft X-ray emission can be identified in the figure, in order of increasing energy, with
K-shell emission of N, O, and Ne, together with several strong Fe L lines. The apparent redshift from the XSTAR emission spectrum fit of
$\sim$0.08$\pm$0.01 was consistent with the EPIC data fits, while - encouragingly -  the RGS data  also requires a steep power law component 
($\Gamma$$\sim$3.3$\pm$0.2).

Figure 8 provides another view of the spectral structure in the soft X-ray band, showing the broad spectral lines that are successfully modelled by
re-emission from the ionised gas in \pg. The best-fit gaussian ($\sigma$$\sim$25 eV at 0.6 keV), would correspond to velocity broadening of 29000 km s$^{-1}$ FWHM, which again would suggest
the high velocity flow occurs over a wide angle. Much better data will be required to determine the true emission line profiles and constrain the
effects of saturation in the principal resonance lines.

\begin{figure}                                                          
\centering                                                              
\includegraphics[width=5.5cm, angle=270]{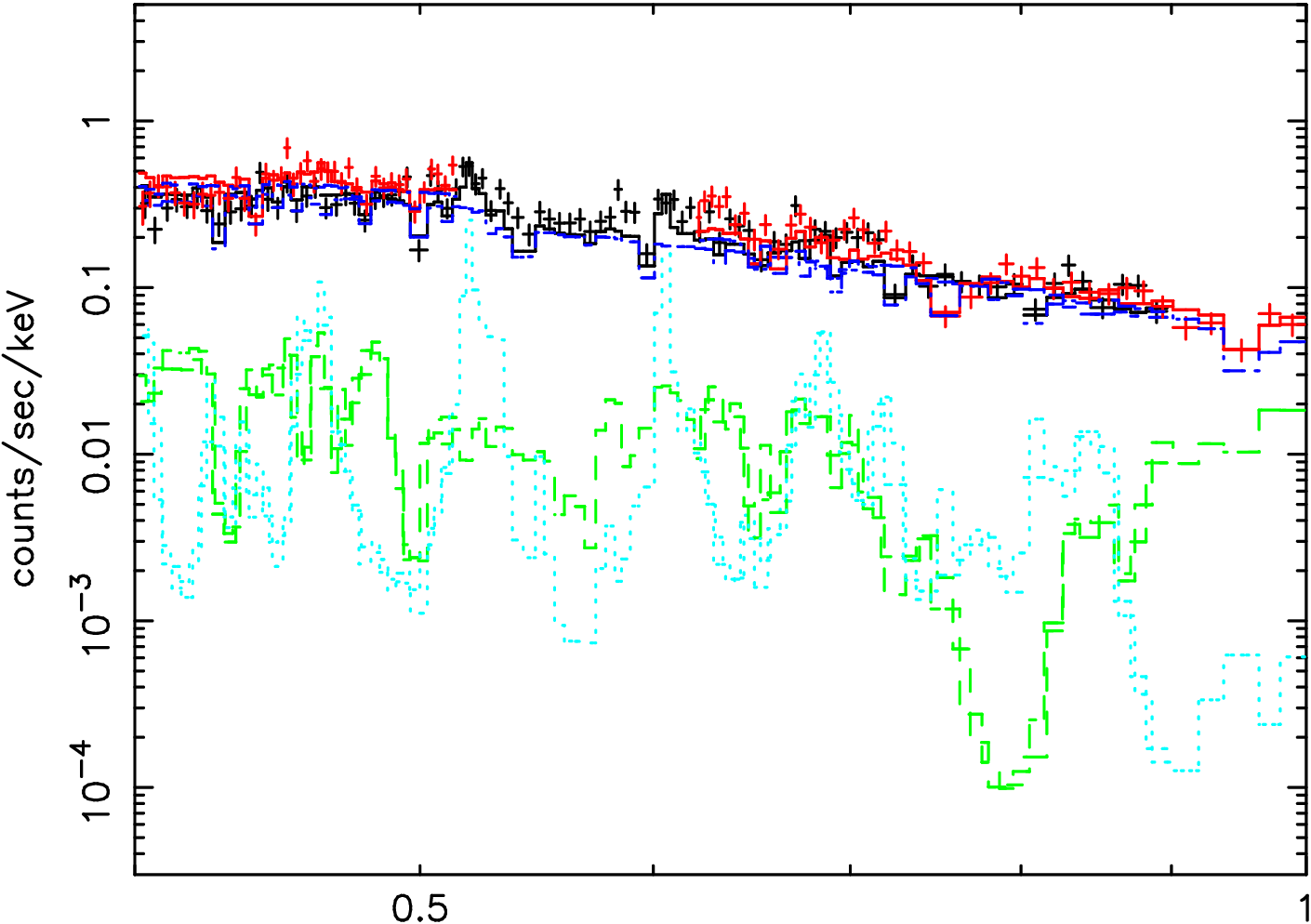}                     
\centering                                                              
\includegraphics[width=5.85cm, angle=270]{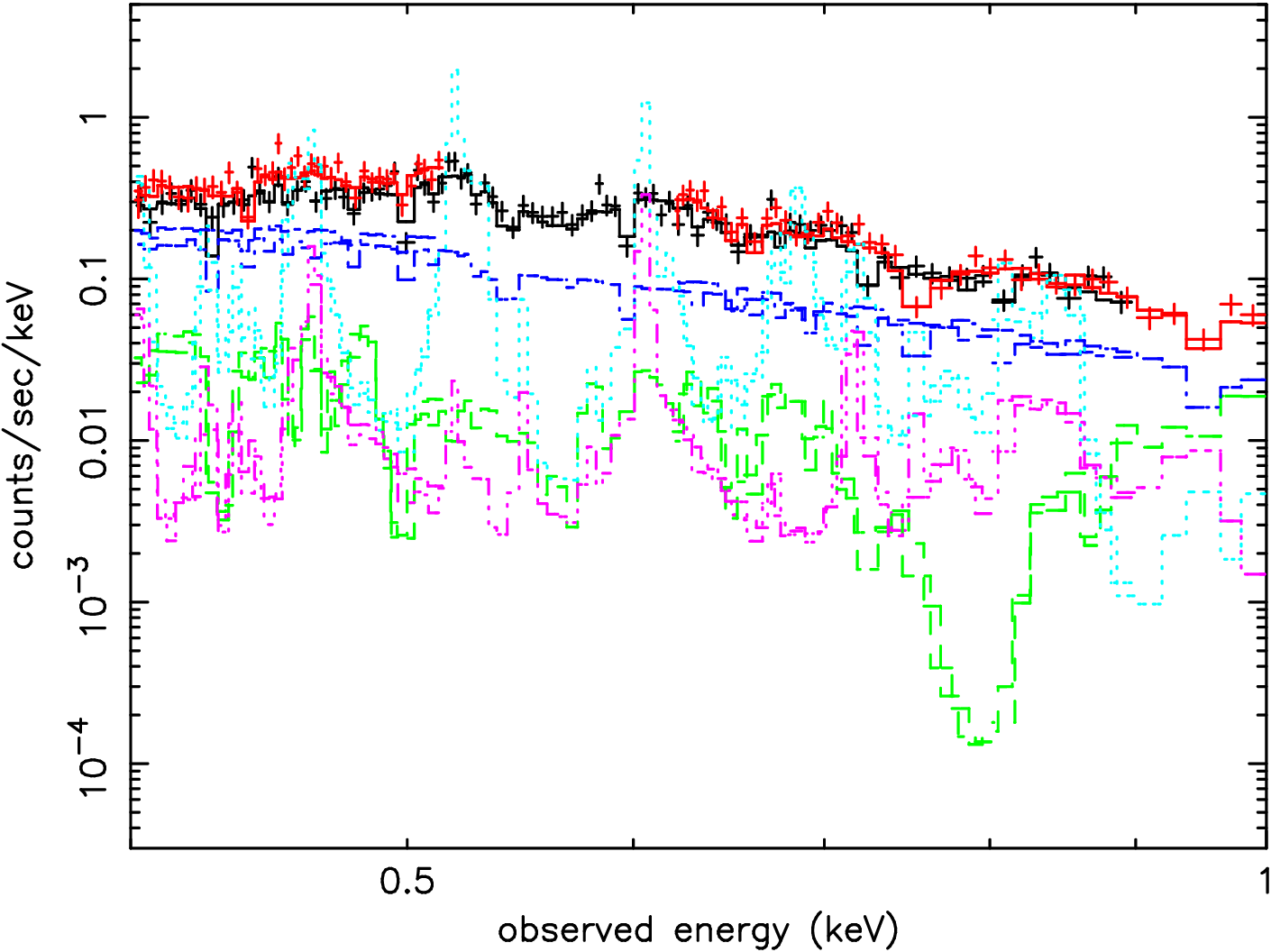}                     
\caption                                                                
{The spectral model fitted to the 2001 RGS data of \pg. The upper panel shows the fit with no added line broadening; the lower panel includes a gaussian 
smoothing of the emission line spectrum providing a much better fit as described in Section 6. The RGS data and overall model fits are shown in red and
black, while the separate model components are the absorbed primary power law (green), secondary power law (dark blue) and input emission line spectra 
for the warm
(light blue) and highly ionised (magenta) gas} 
\end{figure}

\begin{figure}                                                          
\centering                                                              
\includegraphics[width=2.87cm, angle=270]{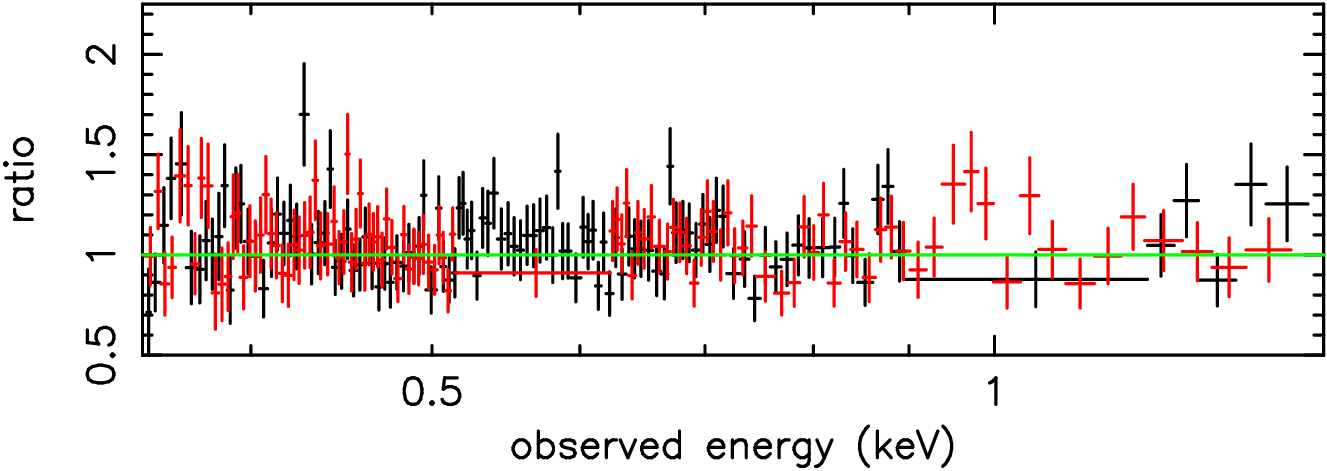}                     
\centering                                                              
\includegraphics[width=4.7cm, angle=270]{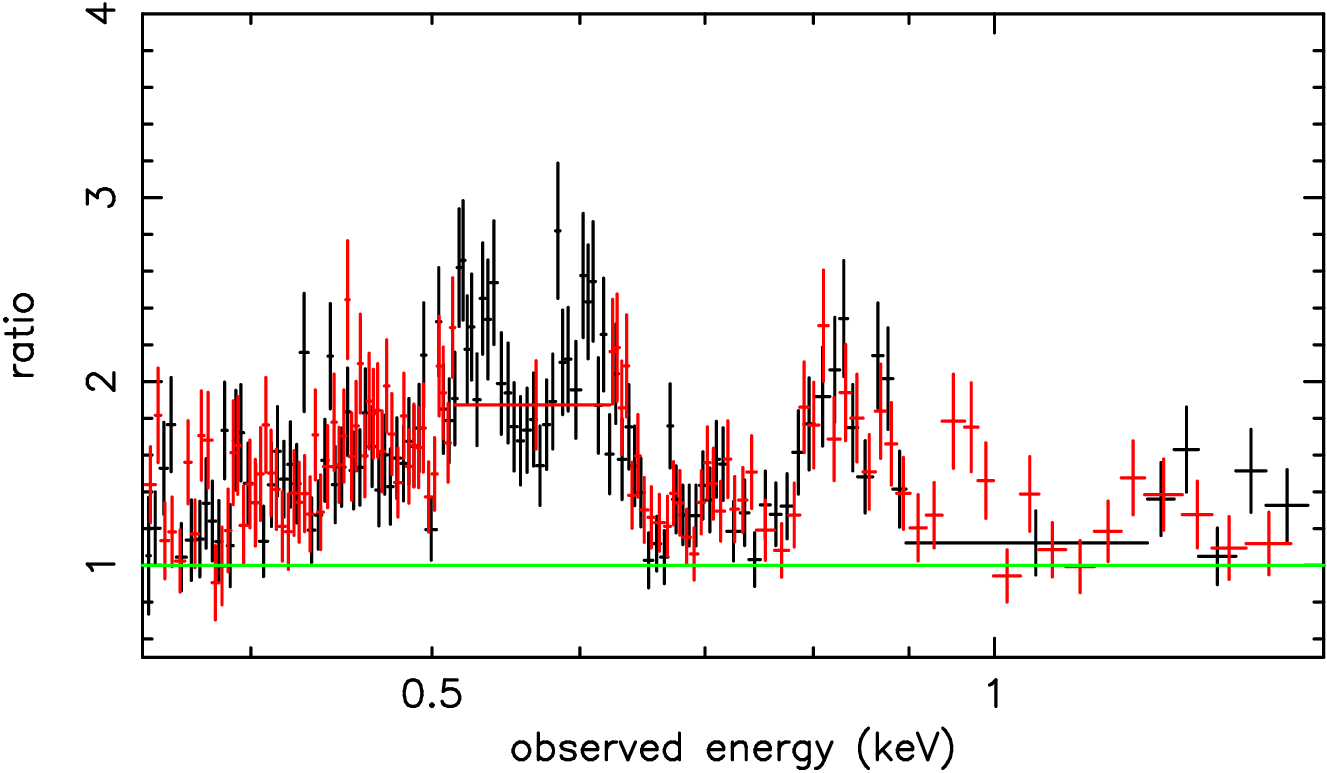}                     
\centering                                                              
\includegraphics[width=4.7cm, angle=270]{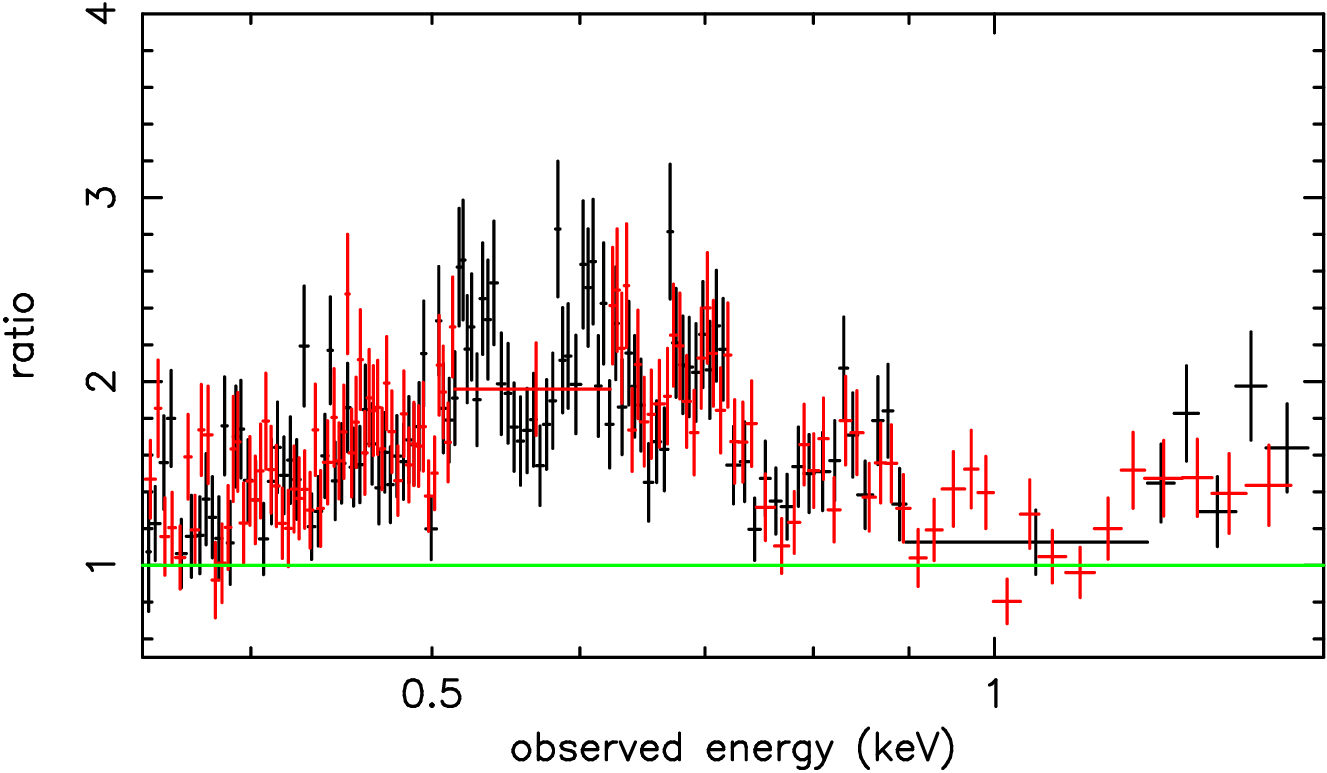}                     
\caption                                                                
{Upper panel) Ratio of the RGS data to the model described in Section 6. The mid panel shows the data:model ratio when the metals C-Mg are removed from the XSTAR 
emission line spectra, revealing 4 broad peaks corresponding (from the left) to resonance line emission from NVII, OVII, OVIII and NeIX. The lower panel shows 
the same data:model ratio when Fe is also removed from the emission line spectra, yielding a further broad excess at $\sim$0.7 keV due to Fe-L emission} 
\end{figure}

The important point for the present study is that the spectral model developed to fit the EPIC data is fully consistent with the RGS data provided the
emission lines are strong but broad. In addition, the RGS spectral fit requires a steep power law continuum, making a
significant contribution to the observed  soft X-ray flux below $\sim$1 keV. The integrated luminosity in the RGS emission line spectrum, of
L$_{em}$$\sim$$1.8\times 10^{43}$~erg s$^{-1}$, is consistent with the values derived from the EPIC spectral fits. To place the soft X-ray emission in a
wider context we note it represents some 0.5\% of the bolometric luminosity of \pg. In comparison, an \xmm\ observation of NGC1068, a Seyfert 2 galaxy
with a similar bolometric luminosity, found an integrated soft X-ray emission luminosity of only $\sim$$5\times 10^{41}$~erg s$^{-1}$ (Pounds and Vaughan
2006). The much stronger soft X-ray line emission from \pg\ supports the view of a strong centrally-condensed emission region that would be
hidden from view in a Seyfert 2.

\section{Applying the new spectral model to the 2004 observation of \pg}

As the new spectral model was based on the evidence for variable ionised  absorption, and re-emission, from a direct comparison of the
2001 and 2004 data sets  it is clearly important to show that the 2001 model parameters can be varied in a  physically reasonable way to also describe the 2004
spectra.  

To that end we tested the new spectral model with the 2004 data of \pg, keeping the element abundances in XSTAR and ionisation parameters unchanged. Justification for
the constant ionisation parameters is based on the near-identical hard X-ray fluxes (which we take as a good proxy for the ionising luminosity). Figure 9
illustrates the fit and the separate model components for the 2004 pn data, which yielded an acceptable $\chi^{2}$  per dof = 691/625. The fit to the 2004
MOS data was statistically excellent, with $\chi^{2}$  per dof = 350/351, as was that to the 2004 RGS data. Comparing the 2001 and 2004 model fits to both
EPIC and RGS data shows that the spectral change is primarily due to a stronger secondary power law component, though the `warm' absorption of the primary
power law and the soft X-ray line emission also appear to be weaker. The column density of the highly ionised absorber is a factor $\sim$2 lower, consistent 
with the significantly weaker
absorption line at $\sim$7 keV in 2004. Dilution by a stronger steep power law continuum and weaker line emission both contribute to
the reduced soft X-ray structure evident in the 2004 RGS spectrum (figure 10).

\begin{figure}                                                          
\centering                                                              
\includegraphics[width=6 cm, angle=270]{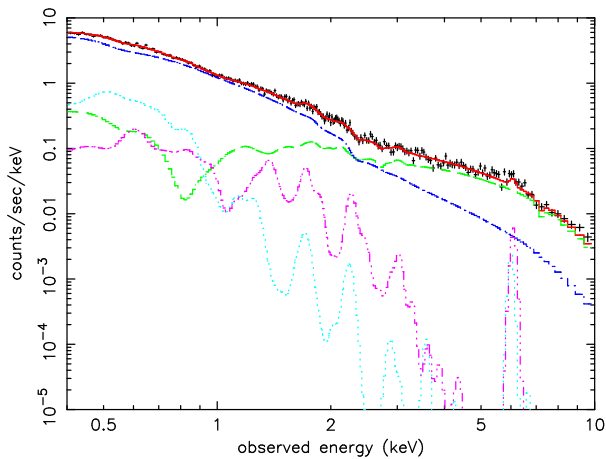}                     
\caption                                                                
{The spectral model fitted to the 2004 pn camera data of \pg. The model components are primary power law (green), additional power law (dark blue), 
re-emission from the moderately ionised gas (light blue) and re-emission from the highly ionised gas (magenta)}
\end{figure}      

\begin{figure}                                                                              
\centering                                                              
\includegraphics[width=4.7cm, angle=270]{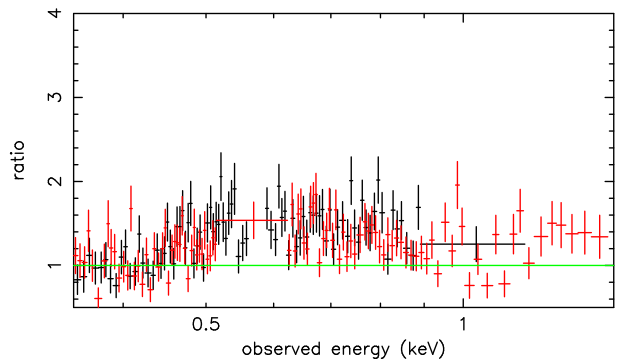}                     
\caption                                                                
{Ratio of the 2004 RGS data to the model shown in figure 9 when the metals C-Fe are removed from the XSTAR 
emission line spectra. Although spectral structure is seen, it is at a significantly lower level than in the corresponding plot for the 2001 RGS data
(figure 8, lower panel)} 
\end{figure}

In summary, we find reasonable changes in the 2001 spectral model also fit the 2004 data. The net effect of a stronger secondary power law, with reduced ionised
absorption (and re-emission) can satisfactorily match the ratio and difference spectra shown in figures 3 and 4. However, given the relative complexity of the 
spectral de-convolution we stress that more than 2 spectral 
`snapshots' will be required to fully resolve and understand the spectral variability. For example, while our modelling suggests the reduced soft X-ray structure
in the 2004 spectrum is a combination of a stronger continuum and weaker line emission, we would expect further observations to show this to be coincidental.

\section{Discussion}

We have explored a new model to describe the complex broad-band X-ray spectrum of the narrow emission line QSO \pg. The model structure was guided 
by a comparison of the spectral data from two \xmm\ observations of \pg\ in 2001 and 2004, which suggested a reduction in ionised absorption (and
re-emission) in the latter case. In addition, the model allows for a second continuum component, described by a steeper power law, to take account of
evidence from other AGN studies that a steep power law can be a principal contributor to spectral variability. Although more complex than the conventional 
description of the X-ray spectrum of \pg, where the hard power law and strong `soft excess' 
are both modelled by Comptonisation of soft accretion disc photons, the new model is encouragingly self-consistent. Thus, the main continuum absorption is
modelled by moderate ionisation (warm) matter of column density $N_{H}$$\sim$$3\times 10^{22}$ cm$^{-2}$ and log$\xi$$\sim$1.4, with re-emission from gas of the
same ionisation parameter dominating the observed spectral structure below $\sim$1 keV. Modelling the higher resolution RGS data confirms the
contributions of the broadened emission lines and the steep power law continuum to the soft X-ray spectrum of \pg. From a comparison of the
absorbed and re-emitted luminosities we find covering factors of $\sim$0.2 and $\sim$0.1, respectively, for the warm and highly ionised outflows. In turn, 
the mechanical energy in the highly ionised outflow is confirmed to be an order of magnitude larger than the luminosity in the secondary power law, supporting our
conjecture that shocks in the outflow could drive that additional continuum component.  

Application of the 2001 spectral model to the 2004 data suggests the primary change is due to an increase in the steep power law continuum rather than 
a decrease in absorption of the primary power
law. In fact, given the quality of the shorter 2004 observation, it is possible the ionised absorption did not change, a possibility consistent with the
very similar ionising fluxes indicated by the comparable hard X-ray spectra which would have required a change to be mainly in column density (or covering factor).
A more extended observation of \pg, exploring spectral variability over hours to days, would be very interesting.

\subsection{The ionised outflow}

The spectral curvature near $\sim$1 keV, previously seen as the onset of a strong `soft excess' in \pg, is attributed in our model primarily to absorption and
re-emission in a substantial column density (N$_{H}$$\sim$$few\times 10^{22}$ cm$^{-2}$) of moderately ionised gas. Although not as well constrained as
the high ionisation gas, which produces an array of narrow absorption lines (Pounds and Page 2006), the warm absorber also appears to be
outflowing at a high velocity. With the present `snapshot' spectra it is not clear how these absorption components might be related;
perhaps they simply represent different matter densities in a common flow.

Identifying the strong absorption line observed at $\sim$7 keV with FeXXV He$\alpha$ (here and in Pounds and Page 2006)
increases the velocity of the highly ionised outflow from $\sim$0.09$\pm$0.01c (P03) to v$\sim$0.14$\pm$0.01c, with a corresponding reduction in the ionisation 
parameter (and
column density) from the XSTAR modelling. These changes can then be applied to the radiatively-driven wind model discussed in P03 and King and Pounds
(2003). 

With an ionising X-ray luminosity ($\ge$7 keV) of 3$\times$$10^{43}$~erg s$^{-1}$ and ionisation parameter $\xi$(=L/nr$^{2}$)$\sim$1000, we have
n$r^{2}$ $\sim$$3\times10^{40}$ cm$^{-1}$. Assuming a spherically symmetric flow, at an outflow velocity  of 0.14c, the mass loss rate is then of order
$\dot M$$ \sim 35b\msun$~yr$^{-1}$, where $b\leq$1 allows for the collimation of the flow. From a comparison of absorbed and re-emitted luminosities in the present
analysis we
found b$\sim$0.1, yielding an outflow mass rate  $\dot M$ $ \sim 3.5\msun$~yr$^{-1}$. This compares with $\dot M_{\rm Edd}$ = 1.6$\msun$~yr$^{-1}$ for a
non-rotating SMBH of mass $\sim$$4\times 10{^7}$$\msun$ (Kaspi \et\ 2000) accreting at an efficiency of 10\%. 

The mechanical energy is now $\sim$$10^{45}$~erg s$^{-1}$, a factor $\sim$15 larger than the luminosity in
the steep power law component. The
ratio of mechanical energy to the bolometric luminosity remains (as in P03) roughly consistent with the value v/c predicted for a radiatively driven outflow (King and
Pounds 2003). Again, 
if the higher velocity equates to the escape velocity at the launch radius R$_{launch}$ (from an optically thick photosphere or
radiatively  extended inner disc), v$\sim$0.14c corresponds to R$_{launch}$ $\sim$ $50R_{\rm s}$  (where $R_{\rm s} = 2GM/c^2$ is the Schwarzschild radius),
or $3\times 10^{14}$~cm.  The EPIC data show significant flux variability in the harder (2-10 keV) band on timescales of 2-3 hours (fig 1 in P03), which would be
compatible with the above scale size relating to the primary (disc/corona) X-ray emission region.

Below $\sim$1 keV we find a similar timescale of variability, but with reduced amplitude. The simplest interpretation is that the photoionised emission is
constant over hours, implying that the secondary power law component is also produced close to the inner disc. Better data will be needed to clarify the time
variability and scale size of the separate model components. 

We have very little information on the geometry of the warm outflow, responsible for most of the continuum absorption (and re-emission), 
except that it presumably originates close to the black hole (from its apparent high velocity) and is largely confined within a radius of $\sim$1 pc (by comparison
with the much weaker soft X-ray emission seen from Seyfert 2s). The observed soft X-ray  luminosity  of L$_{em}$$\sim$$2\times 10^{43}$~erg s$^{-1}$
corresponds to an emission measure $\Sigma$$n^{2}.V$$\sim$$10^{66}$ cm$^{-3}$ for a solar abundance  gas with ionisation parameter $\xi$$\sim$25. With an
ionising luminosity $\sim$$10^{44}$~erg s$^{-1}$, the ionisation parameter gives n.r$^{2}$$\sim$$4\times 10^{42}$. Assuming a spherical shell of radius r,
thickness $\delta$r and density n, a second constraint is the measured column density n.$\delta$r $\sim$ $4\times 10^{22}$ cm$^{-2}$. Interestingly, these
parameters match the above emission measure independent of the assumed scale size. For example, assuming r$\sim$$10^{16}$ cm, we find n$\sim$$4\times
10^{10}$ cm$^{-3}$ and  $\delta$r $\sim$$10^{12}$ cm. For r$\sim$$10^{17}$ cm, an alternative set of self-consistent parameters are n$\sim$$4\times
10^{8}$ cm$^{-3}$ and  $\delta$r $\sim$$10^{14}$ cm.

Although the above estimates are very crude, we note the relatively high densities are essential for much of the gas to remain in a state of moderate ionisation 
so close to the powerful continuum source.
A likely geometry may be for the warm  gas to exist in many, dispersed small clouds rather than the thin spherical shell assumed above. 

That would extend the picture developed in P03 and King and Pounds (2003) to incorporate higher density `clouds' entrapped in the fast outflow and
responsible for the strong low energy absorption (and re-emission) previously interpreted as the `soft excess'. A possible scenario might be where an inhomogeneous
flow
accretes through the inner disc to a radius R where radiation pressure causes the matter to be launched at the local escape velocity. As
the outflow expands outward the mean density will fall, as will the filling factor of the cooler, more opaque, matter. If the secondary power law is indeed
formed by shocks in the fast outflow, then a centrally condensed absorber could explain the stronger absorption we find on the primary power law
continuum (assumed to form in the disc/corona).

\subsection{The second power law component}

The incentive for including a second power law component derived mainly from the detailed studies of MCG-6-30-15 (Vaughan and Fabian
2004) and 1H 0419-577 Pounds \et\ 2004), where difference spectrum analyses showed the primary spectral change in each case could be attributed
to a variable flux, steep
power law. It is also interesting to recall that a `broken power law', with a steeper component at lower energies has often been used to parameterise
AGN spectra, without any physical explanation being offered. In the present study of \pg\ we find the inclusion of a second continuum component 
provides the simplest variable to allow a good fit to both 2001 and 2004 EPIC data. In contrast, a reduction in the covering factor (modelled by a lower column) of
ionised absorber on the primary power law is not able - alone - to fit both data sets.

A further reason to include the second power law component in our modelling was to quantify the conjecture that the the mechanical energy in the high 
velocity outflow in \pg\ might power a continuum component additional to primary disc/coronal emission.  A key factor in determining the mass and energy in
the outflow is the degree of collimation. Our modelling of the broad-band spectrum of \pg\ has  quantified both the absorbed and re-emitted fluxes for the
ionised outflow and thereby allowed the  covering factors (or fractional solid angle) of to be estimated. With a covering factor CF$\sim$0.1, we have
seen that the mechanical energy in the fast, highly ionised outflow is ample to power the second continuum (steep power law) component. Few hours
variability below $\sim$1 keV, although of lower amplitude than in the harder X-ray band, suggests the second continuum component originates at a small
radius. If powered by the fast outflow, possibilities might be internal shocks in the flow (perhaps analogous to the process suggested in Gamma Ray Bursts),
or by running into the slower moving clouds providing the bulk of the continuum opacity. A possible X-ray emission mechanism in either case could be
Comptonisation of optical-UV disc photons, where the steep power law would be a result of the relatively lower energy content of the electron `corona'
compared with that responsible for the primary power law continuum.

\section{A revised assessment of AGN X-ray spectra}

In contemplating how widely applicable our new model might be to describing the X-ray spectra of type 1 AGN, we note that a bolometric luminosity of
order $4\times$$ 10^{45}$~erg   s$^{-1}$ and reverberation mass estimate for the SMBH in \pg\ of $M \sim 4 \times 10^{7}\Msun$ (Kaspi \et\ 2000) indicates
\pg\ is accreting close to the Eddington rate. Additional support for that view comes from  the optical classification of \pg\
as a Narrow Line Seyfert, now thought to be a characteristic of a high accretion ratio (eg Pounds and Vaughan 2000  and references therein). King and
Pounds (2003) provided a simple physical model whereby massive, high velocity outflows can be expected in AGN accreting at or above the Eddington limit.
\pg\ may therefore be a special case among bright, nearby AGN. 

However, it is
intriguing that the X-ray spectra of many bright type 1 AGN have a similar profile to that of \pg\ shown in figure 1, suggesting ionised absorption is a common
cause of the `soft excess'.  An alternative model that works well for weaker soft excesses invokes strong photoionised reflection
(Crummy
\et\ 2006), where the smooth spectral profile is explained by relativistic broadening in the inner disc.   

Further observations of \pg, extended over several days, are essential to test the new spectral model, in particular to resolve the contributions of variable absorption (and
re-emission) from ionised gas to the overall spectral form. The variability timescale of the secondary power law will constrain the scale
of the region where we predict the fast outflow undergoes internal shocks. Detecting hard X-ray emission from \pg\ above $\sim$20 keV would also support the
need for a continuum component with photon index less steep than that in the single power law/ absorption models (Gierlinski and Done 2004, Schurch and
Done 2006). 
Finally, much deeper RGS spectra are needed to resolve the broad emission line profiles indicated in the model fits.    

\section{Summary}

(1) Previous analyses of the 2001 \xmm\ observation of the bright quasar \pg\ have reported evidence of a high velocity ionised outflow (P03, 
Pounds and Page 2006)). By
deconvolving the absorption and emission luminosities in the broad band spectrum, we now find a covering factor of order $\sim$0.1 for the fast outflow,
confirming the mass rate and
mechanical energy to be comparable to the accretion rate and bolometric luminosity, respectively  

(2) We show that an additional outflow component of less highly ionised (and hence more opaque) gas can impose significant curvature  on
the emerging spectrum near $\sim$1 keV, thereby replicating a `soft excess' when applied to a continuum steeper than the canonical value. 

(3) Narrow absorption lines in the EPIC spectra show the outflow to be approximately radial, in contrast with the assumption of a relativistically
smeared absorber in the soft excess modelling of Schurch and Done (2006).

(4) However, like Shurch and Done (2006), we find that re-emission of the absorbed continuum also makes a significant contribution to the observed soft 
X-ray flux. The high integrated
luminosity of the photionised outflow in \pg\ is more than an order of magnitude larger than that of NGC1068, a Seyfert 2 galaxy of 
comparable bolometric luminosity, consistent with an expanding flow originating close to the SMBH. 

(5) Inclusion of a secondary continuum component, which energetically could be powered by the high velocity outflow, is supported by a fit to the soft
X-ray spectrum of the RGS. The RGS analysis also requires the soft X-ray line emission to be strongly broadened, possibly by a combination of velocity
broadening and line saturation. 

(6) Testing the resulting model with data from a second \xmm\ observation of \pg\ in 2004 shows that the spectral change is dominated by an increase
in the secondary power law component, which mimics the effect of reduced ionised absorption of the primary continuum. 

(7) A further interesting consequence of the double power law model is to remove the conceptual difficulty in considering partial covering as an alternative  to
the extreme relativistic Fe K emission line apparent in a simple power law fit (P03).

(8) We suggest that the above model could be generally applicable to type 1 AGN accreting at or above to the Eddington  limit.

\section*{ Acknowledgements }
The results reported here are based on observations obtained with \xmm, an ESA science mission with
instruments and contributions directly funded by ESA Member States and
the USA (NASA).
The authors wish to thank the SOC and SSC teams for organising the \xmm\
observations and initial data reduction.

\end{document}